\documentclass[showpacs,showkeys,preprintnumbers]{revtex4}%
\usepackage{amsfonts}
\usepackage{amsmath}
\usepackage{graphicx}
\usepackage{float}
\usepackage{dcolumn}
\usepackage{bm}
\usepackage{amssymb}%
\begin{document}
\title{Decay of the pseudoscalar glueball and its first excited
state into scalar and pseudoscalar mesons and their first excited
states}
\author{Walaa I.\ Eshraim}
\affiliation{ Institute for Theoretical Physics, Goethe University, Max-von-Laue-Str.\ 1, 60438 Frankfurt am Main, Germany\\
 and Institute for Theoretical Physics, Justus-Liebig University, Heinrich-Buff-Ring 16, 35392 Giessen, Germany}

\begin{abstract}
We expand the study of the pseudoscalar glueball and its first excited state by constructing an interaction Lagrangian which produces the two- and three-body decays of the pseudoscalar glueball, $J^{PC}=0^{-+}$, into the (pseudo)scalar and the excited (pseudo)scalar mesons as well as by constructing other two different chiral Lagrangians which describe the two- and three-body decays of the first excited pseudoscalar glueball, $J^{PC}=0^{*-+}$, into the (pseudo)scalar and the excited (pseudo)scalar mesons. We compute the decay channels for the ground state of a pseudoscalar glueball with a mass of $2.6$ GeV and for the first excited pseudoscalar glueball with a mass $3.7$ GeV, following predictions from lattice QCD in the quenched approximation. These states and channels are within reach of PANDA experiment at the upcoming FAIR facility experiment and ongoing BESIII experiment. In our approach, the various branching ratios are a parameter-free prediction. 

\end{abstract}

\maketitle

\section{Introduction}

Glueballs are predicted as bound states of gluons in models based on quantum chromodynamics (QCD) \cite{bag-glueball}, the theory of fundamental strong interactions of quarks and gluons, or in lattice QCD. The glueball ground state is called scalar glueball, which is estimated to be in the mass range from 1000 to 1800 MeV, followed by a pseudoscalar glueball at higher mass. Due to the non-Abelian nature \cite{Nakano} of the $SU(3)_c$ symmetry, the gauge fields of QCD- the gluons- carry color and interact strongly with themselves, forming colorless states or 'white'. Numerous simulations of lattice QCD confirmed the existence of the bound states, glueball states, and their exotic states to appear in the meson spectrum below 5 GeV \cite{Morningstar, Chen} with different quantum numbers $J^{PC}$. However, the mixing of glueballs ($gg$) and quarkonium ($\overline{q}q$) states, with the same spin, parity and quantum numbers, occurs complicating the experimental search for glueballs because the physical corresponding resonances, which are presented in the Particle Data Group (PDG) \cite{PDG}, emerge producing mixed states. Therefore, there are no glueball states unambiguously identified up to now. Actually, the determination of (predominantly) glueball states is achieved through their decays which should be narrow and exhibit 'flavour blindess'. This makes the present work relevant, as it computes the decays of two different states of glueball, in particular for BESIII experiment \cite{bes}, for the upcoming PANDA experiment at the FAIR facility \cite{panda} and for NICA \cite{NICA} as their program is focussed to establish the existence and the properties of glueballs. \\
\indent In recent years, the properties of glueballs and exotic states has been the focus of many experimental and theoretical hadronic physics studies \cite{bag-glueball, review, Ochs, Sonnenschein} for a deeper understanding of the nonperturbative behavior of QCD. As seen in Refs. \cite{Novikov, Pimikov},  the properties of scalar glueball and its exotic states have been studied in QCD sum rules approach. The hadronic properties of pseudoscalar glueball and its exotic states have been also widely investigated \cite{Masoni, EshraimG, WGNuclion, Volkov, Eshraim-schramm,Brunner} and references therein because they contain an important feature of QCD, the chiral anomaly \cite{Rosenzweig, Ohta}. lattice QCD simulations computed extensively the glueball spectrum \cite{Morningstar, Chen, Lattice}, and predicted the pseudoscalar glueball state, $J^{PC}=0^{-+}$, with a mass of about $2.6$ GeV and the first excited pseudoscalar glueball, $J^{PC}=0^{*-+}$, with a mass of $3.7$ GeV. Both are included in the present investigation. In Ref. \cite{EshraimG}, the branching ratios of the lightest pseudoscalar glueball were computed within a chirally invariant interaction term coupling the pseudoscalar glueball to light mesons. We obtained a dominant channel $\pi\pi K$ and sizable channels, $\pi\pi\eta$ and $\pi\pi\eta'$ which are important for searching the glueball states experimentally. Moreover, in Ref. \cite{WGNuclion}, we computed the decay widths of the lightest pseudoscalar glueball into two nucleons. We studied also the decays of the first excited pseudoscalar glueball by using the same interaction Lagrangian, in the case of $N_f=4$ and found that the excited pseudoscalar glueball decays into the pseudoscalar charmed meson $\eta_c$ as $\Gamma_{\tilde{G}\rightarrow\eta_C \pi\pi}$, and two other chirally invariant terms, one coupling the pseudoscalar glueball with the excited pseudoscalar glueball and (pseudo)scalar mesons as well, and the second coupling the first excited pseudoscalar glueball with a scalar glueball and (pseudo)scalar mesons, as seen in Ref. \cite{Eshraim-schramm}. In the present study, we add the decay channels of a pseudoscalar glueball and its first excited state to excited mesons ($\overline{q}q$), especially excited (pseudo)scalar mesons. The excited scalar and pseudoscalar states correspond to $2 ^3P_0$ and $2 ^1S_0$ configurations, respectively, in spectroscopical notation. Excited mesons have been studied with a wide range of approaches as lattice QCD \cite{ExLattice1, ExLattice2}, QCD string approaches \cite{Badalian}, NJL model \cite{NJL}, Bethe-Salpeter equation \cite{Bethe}, and chiral Lagrangians \cite{excited-mesons, excited-mesons2} as seen recently by the extended linear sigma model (eLSM) \cite{Excited-dick}.\\
\indent The present study of the pseudoscalar glueball and its first excited state is based on the eLSM \cite{Excited-dick}, the effective chiral model of low-energy QCD. The model implements the symmetries of the QCD and their breaking and contains all quark-antiquark mesons with (pseudo)scalar and (axial)vector as well as a scalar and a pseudoscalar glueball. The eLSM played an important role in the study of hadron phenomenology, which has been successfully used to study the vacuum properties of light mesons in the cases of $N_f=2$ \cite{DickNF2}, $N_f=3$ \cite{dick}, glueballs \cite{EshraimG, WGNuclion, Eshraim-schramm, staniglueball}, baryons \cite{Bary}, excited mesons \cite{Excited-dick}, and surprisingly still able to study the vacuum properties of the open and hidden charmed mesons \cite{wcharm1, EshraimTH, wcharm2}.\\
\indent In this work we resume and extend the investigation of the pseudoscalar glueball \cite{EshraimG} and its first excited state \cite{Eshraim-schramm} through their decay channels. We consider now three different chiral Lagrangians describing the two- and three-body decays of the pseudoscalar glueball and its first excited into scalar and pseudoscalar mesons as well as into the excited scalar and pseudoscalar mesons. We obtain within the present approach new channel resonances for both the pseudoscalar glueball and its first excited state which did not appear in Refs. \cite{EshraimG, Eshraim-schramm}. That gives more possibilities for searching glueballs experimentally, by measuring the proposed channels.\\
 This paper is organized as follows. In Sec. II we present the effective Lagrangian interaction between the pseudoscalar glueball with scalar, pseudoscalar, the first excited scalar, and pseudoscalar quark-antiquark degrees of freedom, allowing for the computation of the branching ratios for the decays into $PP,\,PP_E,\, PPS_E$ and $PPP$. In Sec. III we present two chiral Lagrangian terms in the case $N_f=3$: (i) the first couples the first excited pseudoscalar glueball with (pseudo)scalar and the first excited (pseudo)scalar mesons; (ii) the second term interacts the first excited pseudoscalar glueball with the first excited scalar and pseudoscalar mesons. Then, we evaluate the branching ratios for the decays of the first excited pseudoscalar glueball into two- and three-body. Finally, in Sec. IV we present the conclusions.

\section{Decay of the pseudoscalar glueball into (pseudo)scalar and excited (pseudo)scalar mesons}

We introduce a $SU(3)_R\times SU(3)_L$ chiral Lagrangian which couples the pseudoscalar glueball $\tilde{G}\equiv\left\vert gg\right\rangle$
with quantum numbers $J^{PC}=0^{-+}$ to the ordinary (pseudo)scalar and the first excited (pseudo)scalar mesons.
\begin{equation}
\mathcal{L}_{\tilde{G}\Phi\Phi_E}^{int}=c_{\tilde{G}\Phi\Phi_E}\,\tilde{G}\,\left[\left(det\Phi-det\Phi^\dag_E\right)^2+\left(det\Phi^\dag-det\Phi_E\right)^2\right]  \text{ ,}
\label{lag1}%
\end{equation}
where $c_{\tilde{G}\Phi\Phi_E}$ is a dimensionless coupling constant, 
\begin{equation}
\Phi=(S^{a}+iP^{a})t^{a}=\frac{1}{\sqrt{2}}\left(
\begin{array}
[c]{ccc}%
\frac{(\sigma_{N}+a_{0}^{0})+i(\eta_{N}+\pi^{0})}{\sqrt{2}} & a_{0}^{+}%
+i\pi^{+} & K_{S}^{+}+iK^{+}\\
a_{0}^{-}+i\pi^{-} & \frac{(\sigma_{N}-a_{0}^{0})+i(\eta_{N}-\pi^{0})}%
{\sqrt{2}} & K_{S}^{0}+iK^{0}\\
K_{S}^{-}+iK^{-} & \bar{K}_{S}^{0}+i\bar{K}^{0} & \sigma_{S}+i\eta_{S}%
\end{array}
\right)  \; , \label{phimatex}%
\end{equation}
and
\begin{equation}
\Phi_E=(S^{a}_E+iP^{a}_E)t^{a}=\frac{1}{\sqrt{2}}\left(
\begin{array}
[c]{ccc}%
\frac{(\sigma_{NE}+a_{0E}^{0})+i(\eta_{NE}+\pi^{0}_E)}{\sqrt{2}} & a_{0E}^{+}%
+i\pi^{+}_E & K_{SE}^{+}+iK^{+}_E\\
a_{0E}^{-}+i\pi^{-}_E & \frac{(\sigma_{NE}-a_{0E}^{0})+i(\eta_{NE}-\pi^{0}_E)}%
{\sqrt{2}} & K_{SE}^{0}+iK^{0}_E\\
K_{SE}^{-}+iK^{-}_E & \bar{K}_{SE}^{0}+i\bar{K}^{0}_E & \sigma_{SE}+i\eta_{SE}%
\end{array}
\right)  \; , \label{phimatex}%
\end{equation}
are multiplets containing the (pseudo)scalar mesons \cite{dick} and the excited (pseudo)scalar mesons \cite{Excited-dick}, respectively. The $t^a$ are the generators of the group $U(N_f)$.\\
Under $SU_{L}(3)\times SU_{R}(3)$ chiral transformations the multiples $\Phi$ and $\Phi_E$ transform as $\Phi\rightarrow U_{L}\Phi U_{R}^{\dagger}$ and $\Phi_E\rightarrow U_{L}\Phi_E U_{R}^{\dagger}$, respectively, whereas $U_{L(R)}=e^{-i\Theta^a_L(R)^{t^a}}$ are $U(3)_{L(R)}$ matrices, and transform under the charge conjugation $C$ as $\Phi\rightarrow \Phi^T,\,\,\,\,\Phi_E\rightarrow\Phi_E^T$ as well as under the parity $P$ as $\Phi(t,\overrightarrow{x})\rightarrow \Phi^\dag(t,\overrightarrow{x}),\,\,\,\,\Phi_E(t,\overrightarrow{x})\rightarrow \Phi_E^\dag(t,\overrightarrow{x})$, respectively. The determinants of the multiplets $\Phi$ and $\Phi_E$ are invariant under $SU_{L}(3)\times SU_{R}(3)$. However, according to the chiral anomaly, these multiplets are not invariant under the axial $U(1)_A$ transformation.
\begin{equation}
det \Phi \rightarrow det U_A\Phi U_A=e^{-i\Theta^0_A\sqrt{2N_f}}det\Phi\neq det\Phi\,,
\end{equation}
 \begin{equation}
det \Phi_E \rightarrow det U_A\Phi_E U_A=e^{-i\Theta^0_A\sqrt{2N_f}}det\Phi_E\neq det\Phi_E\,.
\end{equation}
On the other hand, the pseudoscalar glueball field $\tilde{G}$ and the excited pseudoscalar field $\tilde{G}^*$ are chirally invariant and transform under the parity $P$ as 
$\tilde{G}(t,\overrightarrow{x})\rightarrow - \tilde{G}(t,\overrightarrow{x}),\,\,\,\,\,\tilde{G}^*(t,\overrightarrow{x})\rightarrow - \tilde{G}^*(t,\overrightarrow{x})\,,$
and under charge conjugation as 
$\tilde{G}\rightarrow \tilde{G},\,\,\,\,\,\tilde{G}^*\rightarrow \tilde{G}^*\,.$ Consequently the effective chiral Lagrangian (\ref{lag1}) contains the symmetries of the QCD Lagrangian. One can see the rest of the mesonic Lagrangian which describes the interactions of $\Phi$ and $\Phi_E$ with a scalar glueball and (axial-)vector degrees of freedom in Sec. A1 of the Appendix and Ref. \cite{Excited-dick} as well.

The scalar and pseudoscalar fields in Eq. (2) are assigned as physical resonances to light quark-antiquark states with mass $\lesssim 2$ GeV \cite{dick}. For the pseudoscalar sector $P$, the fields $\overrightarrow{\pi}$ and $K$ represent the pion isotriplet and the kaon isodoublet respectively \cite{PDG}. The bare quark-antiquark fields $\eta_{N}%
\equiv\left\vert \bar{u}u+\bar{d}d\right\rangle /\sqrt{2}$ and $\eta_{S}%
\equiv\left\vert \bar{s}s\right\rangle $ are the nonstrange and strangeness mixing
components of the physical states $\eta$ and $\eta^{\prime}$ which can be obtained by \cite{PDG}:%
\begin{equation}
\eta=\eta_{N}\cos\varphi+\eta_{S}\sin\varphi,\text{ }\eta^{\prime}=-\eta
_{N}\sin\varphi+\eta_{S}\cos\varphi, \label{mixetas}%
\end{equation}
where the mixing angle is $\varphi\simeq-44.6^{\circ}$ \cite{dick}. For the scalar sector $S$, the field $\vec{a}_{0}$ is assigned to the
physical isotriplet state $a_{0}(1450)$ and the scalar kaon field $K_{S}$ 
to the physical isodoublet state $K_{0}^{\star}(1430).$ In the scalar-isoscalar sector, the nonstrange bare field $\sigma_{N}\equiv\left\vert \bar{u}u+\bar{d}d\right\rangle /\sqrt{2}$
can be assigned to the resonance $f_{0}(1370)$ and the bare strange field $\sigma_{S}$ corresponds to $f_0(1500)$ \cite{staniglueball}, which the two resonances mix with the scalar glueball, G, which refers to $f_{0}(1710)$. The mixing matrix constructed in Ref.\cite{staniglueball} which is given as 
\begin{equation}\label{scalmixmat}
\left(%
\begin{array}{c}
 f_0(1370) \\
 f_0(1500)   \\
 f_0(1710)\\
\end{array}%
\right)=\left(%
\begin{array}{ccccc}
 -0.91 & 0.24 & -0.33\\
 0.30 & 0.94  &-0.17\\
 -0.27 & 0.26 & 0.94\\
\end{array}%
\right)\left(%
\begin{array}{c}
 \sigma_N\\
  \sigma_S\\
  G\\
\end{array}%
\right).
\end{equation}
We now turn to the assignment of the excited states in Eq.(3) as follows: (1) In the excited pseudoscalar sector the excited pion $\overrightarrow{\pi}_E$ and the excited kaon $K_E$ are assigned to $\pi(1300)$ and $K(1460)$, respectively. The excited nonstrange bare fields $\eta_{NE}$ and strange bare field $\eta_{SE}$ correspond to the physical resonances $\eta(1295)$ and $\eta(1440)$, respectively.  
(2) In the excited scalar sector the excited field $\overrightarrow{a}_0$ corresponds to the physical state $a_0(1950)$ and the excited scalar kaon fields $K_{SE}$ is assigned to the resonances $K^*_0(1950)$. The excited scalar-isoscalar sector, the excited nonstrange bare field $\sigma_{NE}\equiv\overline{n}n>$ is identified with the physical resonance $f_0(1790)$ and the excited bare strange field $\sigma_{SE}\equiv\overline{s}s>$ is assigned either to $f_0(2020)$ or to $f_0(2100)$ as has been discussed as a consequence of the model. For more details see Ref. \cite{Excited-dick}. 
To implement the effect of spontaneous symmetry breaking, which takes place, one has to shift the scalar-isoscalar fields by their vacuum expectation values $\phi_N$ and $\phi_S$ as follows \cite{dick} 
\begin{equation}
\sigma_{N}\rightarrow\sigma_{N}+\phi_{N}\text{ and }\sigma_{S}\rightarrow
\sigma_{S}+\phi_{S}\text{ .} \label{shift1}%
\end{equation}
Moreover, when the Lagrangian contains also (axial-)vector mesons, we have to consider the shift of the axial-vector fields and thus redefine the wave-function renormalization constants of the pseudoscalar fields:
\begin{equation}
\vec{\pi}\rightarrow Z_{\pi}\vec{\pi}\text{ , }K^{i}\rightarrow Z_{K}%
K^{i}\text{, }\eta_{j}\rightarrow Z_{\eta_{j}}\eta_{j}\;, \label{psz}%
\end{equation}
where $i=1,2,3$ refers to the four kaonic fields and $j$ refers to $N$ (nonstrange) and $S$ (strange). In Ref. \cite{dick} we find the
numerical values of the renormalization constants of the corresponding wave functions as $Z_{\pi}=1.709$,
$Z_{K}=1.604,Z_{K_{S}}=1.001,$ $Z_{\eta_{N}}=Z_{\pi},$ $Z_{\eta_{S}}=1.539$. The corresponding chiral condensates $\phi_N$ and $\phi_S$ read

\begin{align}
\phi_{N}=&Z_{\pi}f_{\pi}=0.158\text{ GeV, }\,\,\,\,\,\,\,\,\,\phi_{S}=\frac{2Z_{K}f_{K}-\phi_{N}%
}{\sqrt{2}}=0.138\text{ GeV}\;,
\end{align}
where the value of the decay constant of the pion and the kaon are $f_{\pi}=0.0922$ GeV and $f_{K}=0.110$ GeV \cite{PDG}, respectively. One obtains the Lagrangian in Eq.\ (\ref{lag1}) which contains the relevant tree-level vertices for the decay processes of pseudoscalar glueball $\tilde{G}$, see
Appendix (Sec.\ \ref{app2}), after performing the operations in Eqs.\ (\ref{shift1}) and (\ref{psz}).\\
Now we can determine the branching ratios of the pseudoscalar glueball, $\tilde{G}$, for the two- and three-body decay into the excited pseudoscalar $\eta_{NE}$ and the excited scalar $\sigma_{SE}$ and into $\eta$, $\eta'$, $\pi$, $a_0$ and one of the scalar-isoscalar states; $f_0(1370),\,f_0(1500)$, and $f_0(1710)$ which correspond to the scalar glueball \cite{staniglueball}. In order to eliminate the unknown coupling constant, we present the branching ratios relative to the total decay width of the pseudoscalar glueball $\Gamma_{\tilde{G}\Phi\Phi_E}^{tot}$, which are summarized in Table I. (The details of the calculation of the two- and three-body decay are given in Appendix A5.)
\begin{center}%
\begin{table}[H] \centering
\begin{tabular}
[c]{|c|c|c|}\hline
Quantity & The theoretical result \\\hline
$\Gamma_{\tilde{G}\rightarrow \eta\eta}/\Gamma_{\tilde{G}\Phi\Phi_E}^{tot}$ & $0.002$ \\\hline
$\Gamma_{\tilde{G}\rightarrow \eta\eta'}/\Gamma_{\tilde{G}\Phi\Phi_E}^{tot}$ & $0.440$\\\hline
$\Gamma_{\tilde{G}\rightarrow \eta^{\prime}\eta^{\prime}}/\Gamma_{\tilde
{G}\Phi\Phi_E}^{tot}$ & $0.249$ \\\hline
$\Gamma_{\tilde{G}\rightarrow \eta_{SE}\eta}/\Gamma_{\tilde{G}\Phi\Phi_E}^{tot}$ & $0.0085$ \\\hline
$\Gamma_{\tilde{G}\rightarrow \eta_{NE}\eta}/\Gamma_{\tilde{G}\Phi\Phi_E}^{tot}$ & $0.0289$\\\hline
$\Gamma_{\tilde{G}\rightarrow \eta_{NE}\eta^{\prime}}/\Gamma_{\tilde
{G}\Phi\Phi_E}^{tot}$ & $0.2082$ \\\hline
$\Gamma_{\tilde{G}\rightarrow \pi\pi\sigma_{SE}}/\Gamma_{\tilde{G}\Phi\Phi_E}^{tot}$ & $0.00016$ for $\sigma_{SE}\equiv f_0(2020)$\\
&
$0.0000014$   for $\sigma_{SE}\equiv f_0(2100)$ \\\hline
$\Gamma_{\tilde{G}\rightarrow a_0\pi\eta}/\Gamma_{\tilde{G}\Phi\Phi_E}^{tot}$ & $0.0011$ \\\hline
$\Gamma_{\tilde{G}\rightarrow \pi\pi f_0(1370)}/\Gamma_{\tilde{G}\Phi\Phi_E}^{tot}$ & $0.0405$\\\hline
$\Gamma_{\tilde{G}\rightarrow \pi\pi f_0(1500)}/\Gamma_{\tilde{G}\Phi\Phi_E}^{tot}$ & $0.0209$\\\hline
$\Gamma_{\tilde{G}\rightarrow \pi\pi f_0(1710)}/\Gamma_{\tilde{G}\Phi\Phi_E}^{tot}$ & $0.0003$\\\hline
$\Gamma_{\tilde{G}\rightarrow KKf_0(1370)}/\Gamma_{\tilde{G}\Phi\Phi_E}^{tot}$ & $0.00005$ \\\hline
\end{tabular}%
\caption{Branching ratios for the two- and three-body decay of the pseudoscalar glueball $\tilde
{G}$.}%
\end{table}%
\end{center}

\section{Decay of an excited pseudoscalar glueball into scalar-isoscalar, (pseudo)scalar, and excited (pseudo)scalar states}

We consider a $SU(3)_R\times SU(3)_L$ chiral Lagrangian that couples the excited pseudoscalar glueball $\tilde{G}^*\equiv\left\vert gg\right\rangle$ with quantum numbers $J^{PC}=0^{-+*}$ to (pseudo)scalar and excited (pseudo)scalar mesons by the same means as the coupling of the pseudoscalar glueball to (pseudo)scalar and excited (pseudo)scalar quark-antiquark states as seen in Eq. (\ref{lag1})
\begin{equation}
\mathcal{L}_{\tilde{G}^*\Phi\Phi_E}^{int}=c_{\tilde{G}^*\Phi\Phi_E}\,\tilde{G}^*\,\left[\left(det\Phi-det\Phi^\dag_E\right)^2+\left(det\Phi^\dag-det\Phi_E\right)^2\right]  \text{ ,}
\label{intlag3}%
\end{equation}
where $c_{\tilde{G}^*\Phi\Phi_E}$ is a dimensionless coupling constant. The effective chiral Lagrangian of Eq. (\ref{intlag3}) is also invariant under $SU_{L}(3)\times SU_{R}(3)$ and parity and realizes the symmetries of the QCD Lagrangian. By using Eqs. (\ref{shift1}) and (\ref{psz}), we get the Lagrangian in Eq. (\ref{intlag2}), which involves the relevant tree-level vertices for the decay processes of the excited pseudoscalar glueball $\tilde{G}^*$, see Appendix (Sec. A 4).\\
In Tables II and III, we present the results of the branching ratios of the excited pseudoscalar glueball $\tilde{G}^*$ for two-body decay widths into (pseudo)scalar, excited (pseudo)scalar mesons, and scalar-isoscalar states, $f_0(1370),\,f_0(1500)$ and $f_0(1710)$, by including the full mixing pattern above $1$ GeV, where the resonance $f_0(1710)$ \cite{staniglueball} corresponds to a scalar meson.

\begin{center}%
\begin{table}[H] \centering
\begin{tabular}
[c]{|c|c|c|c|c|c|c|}\hline
 Case (i):$\mathcal{L}_{\tilde{G}^*\Phi\Phi_E}^{int}$ & The theoretical result \\\hline
$\Gamma_{\tilde{G}^*\rightarrow \eta\eta}/\Gamma_{\tilde{Ge}^*\Phi\Phi_E}^{tot}$ & $7.399\times10^{-7}$ \\\hline
$\Gamma_{\tilde{G}^*\rightarrow \eta\eta'}/\Gamma_{\tilde{G}^*\Phi\Phi_E}^{tot}$ & $1.9\times10^{-4}$\\\hline
$\Gamma_{\tilde{G}^*\rightarrow \eta^{\prime}\eta^{\prime}}/\Gamma_{\tilde
{G}\Phi\Phi_E}^{tot}$ & $1.3\times10^{-4}$ \\\hline
$\Gamma_{\tilde{G}^*\rightarrow \eta_{NE}\eta_{NE}}/\Gamma_{\tilde{G}^*\Phi\Phi_E}^{tot}$ & $6.8\times10^{-5}$ \\\hline
$\Gamma_{\tilde{G}^*\rightarrow \eta_{SE}\eta_{SE}}/\Gamma_{\tilde{G}^*\Phi\Phi_E}^{tot}$ & $7.16\times 10^{-6}$\\\hline
$\Gamma_{\tilde{G}^*\rightarrow \eta_{SE}\eta}/\Gamma_{\tilde
{G}^*\Phi\Phi_E}^{tot}$ & $4.13\times 10^{-6}$ \\\hline
$\Gamma_{\tilde{G}^*\rightarrow \eta_{SE}\eta_{NE}}/\Gamma_{\tilde{G}^*\Phi\Phi_E}^{tot}$ & $4.4\times10^{-5}$ \\\hline
$\Gamma_{\tilde{G}^*\rightarrow \eta_{SE}\eta'}/\Gamma_{\tilde{G}^*\Phi\Phi_E}^{tot}$ & $6.2\times10^{-5}$\\\hline
$\Gamma_{\tilde{G}^*\rightarrow \eta_{NE}\eta}/\Gamma_{\tilde
{G}^*\Phi\Phi_E}^{tot}$ & $1.5\times10^{-5}$ \\\hline
$\Gamma_{\tilde{G}^*\rightarrow \eta_{NE}\eta'}/\Gamma_{\tilde{G}^*\Phi\Phi_E}^{tot}$ & $1.9\times10^{-4}$ \\\hline

$\Gamma_{\tilde{G}^*\rightarrow \sigma_{NE}\sigma_ {NE}}/\Gamma_{\tilde{G}^*\Phi\Phi_E}^{tot}$ & $1.1\times10^{-5}$ \\\hline
\end{tabular}%
\caption{Branching ratios for the two-body decay of the excited pseudoscalar glueball $\tilde
{G}^*$ into the (pseudo)scalar and excited (pseudo)scalar mesons.}%
\end{table}%
\end{center}

\begin{center}%
\begin{table}[H] \centering
\begin{tabular}
[c]{|c|c|c|c|c|c|c|}\hline
 Case (i):$\mathcal{L}_{\tilde{G}^*\Phi\Phi_E}^{int}$ & The theoretical result \\\hline
$\Gamma_{\tilde{G}^*\rightarrow f_0(1370)\sigma_{NE}}/\Gamma_{\tilde{G}^*\Phi\Phi_E}^{tot}$ & $0.3\times10^{-4}$ \\\hline
$\Gamma_{\tilde{G}^*\rightarrow f_0(1500)\sigma_{NE}}/\Gamma_{\tilde{G}^*\Phi\Phi_E}^{tot}$ & $2.2\times10^{-5}$ \\\hline
$\Gamma_{\tilde{G}^*\rightarrow f_0(1710)\sigma_{NE}}/\Gamma_{\tilde{G}^*\Phi\Phi_E}^{tot}$ & $6.54\times10^{-7}$ \\\hline
$\Gamma_{\tilde{G}^*\rightarrow f_0(1370)\sigma_{SE}}/\Gamma_{\tilde{G}^*\Phi\Phi_E}^{tot}$ & $8.2\times 10^{-6}$    for $\sigma_{SE}\equiv f_0(2020)$\\
&
$6.77\times 10^{-6}$   for $\sigma_{SE}\equiv f_0(2100)$
\\\hline
$\Gamma_{\tilde{G}^*\rightarrow f_0(1500)\sigma_{SE}}/\Gamma_{\tilde{G}^*\Phi\Phi_E}^{tot}$ & $9.18\times 10^{-8}$  for $\sigma_{SE}\equiv f_0(2020)$\\
&
$6.26\times 10^{-8}$   for $\sigma_{SE}\equiv f_0(2100)$
\\\hline
$\Gamma_{\tilde{G}^*\rightarrow f_0(1370)f_0(1370)}/\Gamma_{\tilde{G}^*\Phi\Phi_E}^{tot}$ & $1.2\times10^{-5}$\\\hline
$\Gamma_{\tilde{G}^*\rightarrow f_(1500) f_0(1500)}/\Gamma_{\tilde{G}^*\Phi\Phi_E}^{tot}$ & $7.99\times 10^{-7}$\\\hline
$\Gamma_{\tilde{G}^*\rightarrow f_(1710) f_0(1710)}/\Gamma_{\tilde{G}^*\Phi\Phi_E}^{tot}$ & $2.81\times 10^{-10}$\\\hline
$\Gamma_{\tilde{G}^*\rightarrow f_0(1370)f_0(1500)}/\Gamma_{\tilde{G}^*\Phi\Phi_E}^{tot}$ & $2.7\times10^{-5}$ \\\hline
$\Gamma_{\tilde{G}^*\rightarrow f_0(1370)f_0(1710)}/\Gamma_{\tilde{G}^*\Phi\Phi_E}^{tot}$ & $4.71\times 10^{-7}$\\\hline
$\Gamma_{\tilde{G}^*\rightarrow f_0(1500)f_0(1710)}/\Gamma_{\tilde{G}^*\Phi\Phi_E}^{tot}$ & $1.99\times 10^{-6}$  \\\hline
\end{tabular}%
\caption{Branching ratios for the two-body decay of the excited pseudoscalar glueball $\tilde
{G}^*$ into the scalar-isoscalar states and excited (pseudo)scalar mesons.}%
\end{table}%
\end{center}

In the following Tables IV and V, we list the results for the branching ratios of $\tilde{G}^*$ of the three-body decay widths into (pseudo)scalar, excited (pseudo)scalar mesons and scalar-isoscalar states, $f_0(1370),\,f_0(1500)$, and $f_0(1710)$ which correspond to a scalar glueball.
\begin{center}%
\begin{table}[H] \centering
\begin{tabular}
[c]{|c|c|c|c|c|c|c|}\hline
 Case (i):$\mathcal{L}_{\tilde{G}^*\Phi\Phi_E}^{int}$ & The theoretical result \\\hline
$\Gamma_{\tilde{G}^*\rightarrow a_0\pi\eta}/\Gamma_{\tilde{G}^*\Phi\Phi_E}^{tot}$ & $2.2\times10^{-5}$ \\\hline
$\Gamma_{\tilde{G}^*\rightarrow a_0\pi\eta'}/\Gamma_{\tilde{G}^*\Phi\Phi_E}$ & $1.7\times10^{-4}$ \\\hline
$\Gamma_{\tilde{G}^*\rightarrow a_0\pi\eta_{NE}}/\Gamma_{\tilde{G}^*\Phi\Phi_E}^{tot}$ & $6.5\times10^{-5}$ \\\hline
$\Gamma_{\tilde{G}^*\rightarrow a_0\pi\eta_{SE}}/\Gamma_{\tilde{G}^*\Phi\Phi_E}$ & $1.5\times10^{-5}$ \\\hline
$\Gamma_{\tilde{G}^*\rightarrow K K_S \eta}/\Gamma_{\tilde{G}^*\Phi\Phi_E}^{tot}$ & $9.3\times10^{-5}$ \\\hline
$\Gamma_{\tilde{G}^*\rightarrow K K_S \eta'}/\Gamma_{\tilde{G}^*\Phi\Phi_E}^{tot}$ & $4.4\times10^{-5}$ \\\hline
$\Gamma_{\tilde{G}^*\rightarrow K K_S \eta_{NE}}/\Gamma_{\tilde{G}^*\Phi\Phi_E}^{tot}$ & $3.2\times10^{-5}$ \\\hline
$\Gamma_{\tilde{G}^*\rightarrow K K \sigma_{NE}}/\Gamma_{\tilde{G}^*\Phi\Phi_E}^{tot}$ &$8.7\times10^{-5}$ \\\hline
$\Gamma_{\tilde{G}^*\rightarrow \pi\pi \sigma_{SE}}/\Gamma_{\tilde{G}^*\Phi\Phi_E}^{tot}$ & $1.1\times10^{-5}$  for $\sigma_{SE}\equiv f_0(2020)$\\
&
$9.5\times10^{-6}$   for $\sigma_{SE}\equiv f_0(2100)$ \\\hline
$\Gamma_{\tilde{G}^*\rightarrow \eta\eta\sigma_{NE}}/\Gamma_{\tilde{G}^*\Phi\Phi_E}^{tot}$ &$0.998$ \\\hline
$\Gamma_{\tilde{G}^*\rightarrow \eta\eta'\sigma_{NE}}/\Gamma_{\tilde{G}^*\Phi\Phi_E}^{tot}$ &$0.8\times10^{-4}$ \\\hline
$\Gamma_{\tilde{G}^*\rightarrow \eta\eta\sigma_{SE}}/\Gamma_{\tilde{G}^*\Phi\Phi_E}^{tot}$ &$0.4\times10^{-6}$ for $\sigma_{SE}\equiv f_0(2020)$\\
&
$2.7\times10^{-7}$   for $\sigma_{SE}\equiv f_0(2100)$ \\\hline
$\Gamma_{\tilde{G}^*\rightarrow \eta\eta'\sigma_{SE}}/\Gamma_{\tilde{G}^*\Phi\Phi_E}^{tot}$ &$1.9\times10^{-7}$ for $\sigma_{SE}\equiv f_0(2020)$\\
&
$3.9\times10^{-7}$ for $\sigma_{SE}\equiv f_0(2100)$ \\\hline
$\Gamma_{\tilde{G}^*\rightarrow \eta\eta_{NE}\sigma_{NE}}/\Gamma_{\tilde{G}^*\Phi\Phi_E}^{tot}$ &$9.9\times10^{-7}$\\\hline
\end{tabular}%
\caption{Branching ratios for the three-body decay of the excited pseudoscalar glueball $\tilde
{G}^*$ into the (pseudo)scalar and excited (pseudo)scalar mesons.}%
\end{table}%
\end{center}

\begin{center}%
\begin{table}[H] \centering
\begin{tabular}
[c]{|c|c|c|c|c|c|c|}\hline
 Case (i):$\mathcal{L}_{\tilde{G}^*\Phi\Phi_E}^{int}$ & The theoretical result \\\hline
$\Gamma_{\tilde{G}^*\rightarrow K K_S f_0(13700)}/\Gamma_{\tilde{G}^*\Phi\Phi_E}^{tot}$ &$8.1\times10^{-5}$ \\\hline
$\Gamma_{\tilde{G}^*\rightarrow K K_S f_0(1500)}/\Gamma_{\tilde{G}^*\Phi\Phi_E}^{tot}$ &$5.8\times10^{-5}$ \\\hline
$\Gamma_{\tilde{G}^*\rightarrow K K_S f_0(1710)}/\Gamma_{\tilde{G}^*\Phi\Phi_E}^{tot}$ &$1.88\times 10^{-6}$ \\\hline
$\Gamma_{\tilde{G}^*\rightarrow K K f_0(13700)}/\Gamma_{\tilde{G}^*\Phi\Phi_E}^{tot}$ &$7.9\times10^{-5}$ \\\hline
$\Gamma_{\tilde{G}^*\rightarrow K K f_0(1500)}/\Gamma_{\tilde{G}^*\Phi\Phi_E}^{tot}$ &$7.22\times 10^{-6}$ \\\hline
$\Gamma_{\tilde{G}^*\rightarrow \eta\eta f_0(13700)}/\Gamma_{\tilde{G}^*\Phi\Phi_E}^{tot}$ &$1.5\times10^{-4}$ \\\hline
$\Gamma_{\tilde{G}^*\rightarrow \eta\eta f_0(1500)}/\Gamma_{\tilde{G}^*\Phi\Phi_E}^{tot}$ &$2.1\times10^{-5}$ \\\hline
$\Gamma_{\tilde{G}^*\rightarrow \eta\eta f_0(1710)}/\Gamma_{\tilde{G}^*\Phi\Phi_E}^{tot}$ &$1.4\times10^{-5}$ \\\hline
$\Gamma_{\tilde{G}^*\rightarrow \eta\eta' f_0(13700)}/\Gamma_{\tilde{G}^*\Phi\Phi_E}^{tot}$ &$7.9\times10^{-5}$ \\\hline
$\Gamma_{\tilde{G}^*\rightarrow \eta\eta' f_0(1500)}/\Gamma_{\tilde{G}^*\Phi\Phi_E}^{tot}$ &$1.02\times10^{-4}$ \\\hline
$\Gamma_{\tilde{G}^*\rightarrow \eta\eta' f_0(1710)}/\Gamma_{\tilde{G}^*\Phi\Phi_E}^{tot}$ &$3.02\times 10^{-7}$ \\\hline
$\Gamma_{\tilde{G}^*\rightarrow \eta'\eta' f_0(13700)}/\Gamma_{\tilde{G}^*\Phi\Phi_E}^{tot}$ &$1.2\times10^{-4}$ \\\hline
$\Gamma_{\tilde{G}^*\rightarrow \eta'\eta' f_0(1500)}/\Gamma_{\tilde{G}^*\Phi\Phi_E}^{tot}$ &$2.3\times10^{-5}$ \\\hline
$\Gamma_{\tilde{G}^*\rightarrow \eta'\eta' f_0(1710)}/\Gamma_{\tilde{G}^*\Phi\Phi_E}^{tot}$ &$8.45\times 10^{-8}$ \\\hline
$\Gamma_{\tilde{G}^*\rightarrow \eta'\eta_{NE} f_0(13700)}/\Gamma_{\tilde{G}^*\Phi\Phi_E}^{tot}$ &$1.02\times 10^{-6}$ \\\hline
$\Gamma_{\tilde{G}^*\rightarrow \eta\eta_{NE} f_0(13700)}/\Gamma_{\tilde{G}^*\Phi\Phi_E}^{tot}$ &$1.42\times 10^{-6}$ \\\hline
$\Gamma_{\tilde{G}^*\rightarrow \eta\eta_{NE} f_0(1500)}/\Gamma_{\tilde{G}^*\Phi\Phi_E}^{tot}$ &$3.84\times 10^{-6}$ \\\hline
$\Gamma_{\tilde{G}^*\rightarrow \eta\eta_{NE} f_0(1710)}/\Gamma_{\tilde{G}^*\Phi\Phi_E}^{tot}$ &$6.72\times 10^{-8}$ \\\hline
$\Gamma_{\tilde{G}^*\rightarrow \eta\eta_{SE} f_0(13700)}/\Gamma_{\tilde{G}^*\Phi\Phi_E}^{tot}$ &$1.66\times 10^{-6}$ \\\hline
$\Gamma_{\tilde{G}^*\rightarrow \eta\eta_{SE} f_0(1500)}/\Gamma_{\tilde{G}^*\Phi\Phi_E}^{tot}$ &$7.09\times 10^{-8}$ \\\hline
$\Gamma_{\tilde{G}^*\rightarrow \eta\eta_{SE} f_0(1710)}/\Gamma_{\tilde{G}^*\Phi\Phi_E}^{tot}$ &$2.5\times 10^{-14}$ \\\hline
$\Gamma_{\tilde{G}^*\rightarrow KK f_0(13700)}/\Gamma_{\tilde{G}^*\Phi\Phi_E}^{tot}$ &$7.9\times10^{-5}$ \\\hline
$\Gamma_{\tilde{G}^*\rightarrow KK f_0(1500)}/\Gamma_{\tilde{G}^*\Phi\Phi_E}^{tot}$ &$7.22\times 10^{-6}$ \\\hline
$\Gamma_{\tilde{G}^*\rightarrow KK f_0(1710)}/\Gamma_{\tilde{G}^*\Phi\Phi_E}^{tot}$ &$3.66\times 10^{-6}$ \\\hline
\end{tabular}%
\caption{Branching ratios for the three-body decay of the excited pseudoscalar glueball $\tilde
{G}^*$ into the scalar-isoscalar states, (pseudo)scalar and excited (pseudo)scalar mesons.}%
\end{table}%
\end{center}

As a second step, we consider the effective chiral Lagrangian that couples the excited pseudoscalar glueball field, $\tilde{G}^*$ to the excited scalar and pseudoscalar mesons.

\begin{equation}
\mathcal{L}_{\tilde{G}^*\phi_E}^{int}=ic_{\tilde{G}^*\Phi_E}\tilde{G}^*\left(
\text{\textrm{det}}\Phi_E-\text{\textrm{det}}\Phi^{\dag}_E\right)  \text{ ,}
\label{intlag2}%
\end{equation}

where $c_{\tilde{G}^*\Phi_E}$ is an unknown coupling constant and $\Phi_E$ is a multiplet of excited scalar and pseudoscalar mesons in the case of $N_f=3$ as shown in Eq.(\ref{phimatex}). The effective Lagrangian of Eq.(\ref{intlag2}) is invariant under $SU_{L}(3)\times SU_{R}(3)$ and parity and fulfills the symmetries of the QCD Lagrangian. \\
Once the operations in Eqs. (\ref{shift1}) and (\ref{psz}) have been performed, the Lagrangian in Eq. (\ref{intlag2}) includes the relevant tree-level vertices for the decay processes of the excited pseudoscalar glueball $\tilde{G}^*$, see Appendix (Sec. A 3). We compute the branching ratios of the two-body decay for the excited pseudoscalar glueball into excited scalar-pseudoscalar mesons relative to the total decay width of the excited pseudoscalar glueball $\Gamma_{\tilde{G}^*\Phi_E}^{tot}$, the results of which are listed in Table VI. \\
\begin{center}%
\begin{table}[H] \centering
\begin{tabular}
[c]{|c|c|c|c|c|}\hline
 Quantity & Case(i): $\sigma_{SE}\equiv f_0(2020)$ & Case (ii): $\sigma_{SE}\equiv f_0(2100)$ \\\hline
 $\Gamma_{\tilde{G}^*\rightarrow a_{0E}\pi_E}/\Gamma_{\tilde{G}^*\Phi_E}^{tot}$ & $0.367$ & $0.375$\\\hline
 $\Gamma_{\tilde{G}^*\rightarrow K_{E}K_{SE}}/\Gamma_{\tilde{G}^*\Phi_E}^{tot}$ & $0.223$ & $0.227$ \\\hline
 $\Gamma_{\tilde{G}^*\rightarrow \eta_{NE} \sigma_{NE}}/\Gamma_{\tilde{G}^*\Phi_E}^{tot}$ & $0.105$ & $0.107$  \\\hline 
 $\Gamma_{\tilde{G}^*\rightarrow \eta_{NE} \sigma_{SE}}/\Gamma_{\tilde{G}^*\Phi_E}^{tot}$ & $0.147$ & $0.129$ \\\hline
 $\Gamma_{\tilde{G}^*\rightarrow \eta_{SE} \sigma_{NE}}/\Gamma_{\tilde{G}^*\Phi_E}^{tot}$  & $0.159$ & $0.162$ \\\hline
\end{tabular}%
\caption{Branching ratios for the two-body decays of the excited pseudoscalar glueball $\tilde
{G}^*$ into the excited (pseudo)scalar mesons}%
\end{table}%
\end{center}

\section{Conclusion}

In this work we have presented three chirally invariant terms, for the three flavor case $N_f=3$, describing two- and three-body decays of a pseudoscalar glueball and a first excited pseudoscalar glueball into scalar and pseudoscalar mesons as well as excited scalar and pseudoscalar mesons. In the first Lagrangian, the decay channels of the pseudoscalar glueball into two-body ($PP, PP_E$) and three-body ($PPS_E, PPS$) which include the scalar-isoscalar states have been computed. We have computed from the second effective Lagrangian the decays of the excited pseudoscalar glueball into two and three (pseudo)scalar mesons, excited (pseudo)scalar mesons and scalar-isoscalar states $f_0(1370)$, $f_0(1500)$ and $f_0(1710)$, where the resonance $f_0(1710)$ corresponds to the scalar glueball. The third interaction Lagrangian produces the decay widths of the excited pseudoscalar glueball into two excited (pseudo)scalar mesons as seen in Table VI. In agreement with lattice QCD in the quenched approximation, we have chosen the mass of the pseudoscalar glueball $2.6$ GeV and the mass of the excited pseudoscalar glueball $3.7$ GeV. While the coupling constant cannot be determined, we predict the results as branching ratios that thus determine the expectation of the dominant decay channels. The existence and the decay properties of the pseudoscalar glueball and its excitations represent a useful guideline for the corresponding upcoming experiments with the PANDA detector at FAIR, for BESIII experiment and for NICA. So, our approach is very interesting for the search of the pseudoscalar glueball and its excitations.\\
\indent In the future, one can see that when lattice QCD works beyond the quenched approximation and include the effect of dynamic fermions, it obtains new results for the pseudoscalar glueball and its excitations which would be very useful for our models.

\section*{Acknowledgements}

The author thanks C. S. Fischer, D. H. Rischke and E. Tomasi for useful discussions. Financial support from the BMBF under Contract No. 05H15RGKBA and HIC for FAIR. 

\appendix

\section{Details of the calculation}

\subsection{The full mesonic Lagrangian}

\label{app1}

The chirally invariant $U(N_{f})_{L}\times U(N_{f})_{R}$ Lagrangian for the excited (pseudo)scalar, (pseudo)scalar and (axial)vector quarkonia with terms up to order four in the naive scaling has the form

\begin{align}
\mathcal{L}_{mesE}  &  =\mathrm{Tr}[(D_{\mu}\Phi_E)^{\dagger}(D_{\mu}\Phi_E
)]+\alpha \mathrm{Tr}[(D_{\mu}\Phi_E)^{\dagger}(D_{\mu}\Phi
)+(D_{\mu}\Phi)^{\dagger}(D_{\mu}\Phi_E
)]
-(m_{0}^*)^{2}\mathrm{Tr}\left(\frac{G}{G_0}\right)^2(\Phi_E^{\dagger}\Phi_E)\nonumber\\&
-\lambda_{0}\left(\frac{G}{G_0}\right)^2\mathrm{Tr}%
(\Phi_E^{\dagger}\Phi+\Phi^\dagger \Phi_E)-\lambda_{1}^*\mathrm{Tr}%
(\Phi^{\dagger}_E\Phi_E)\mathrm{Tr}(\Phi^{\dagger}\Phi
)-\lambda_{2}^*\mathrm{Tr}(\Phi_E^{\dagger}\Phi_E\Phi^\dagger\Phi+\Phi_E\Phi^\dagger_E\Phi\Phi^\dagger)\nonumber\\&
 -\kappa_1\mathrm{Tr}%
(\Phi_E^{\dagger}\Phi+\Phi^\dagger \Phi_E)\mathrm{Tr}(\Phi^{\dagger}\Phi
)-\kappa_2[\mathrm{Tr}%
(\Phi_E^{\dagger}\Phi+\Phi^\dagger \Phi_E)]^2-\kappa_3(\Phi_E^{\dagger}\Phi+\Phi^\dagger \Phi_E)\mathrm{Tr}(\Phi^{\dagger}_E\Phi_E
)-\kappa_4[\mathrm{Tr}(\Phi^{\dagger}_E\Phi_E
)]^2\nonumber\\& -\xi_1\mathrm{Tr}(\Phi_E^{\dagger}\Phi\Phi^\dagger\Phi+\Phi_E\Phi^\dagger\Phi\Phi^\dagger)-\xi_2\mathrm{Tr}(\Phi_E^{\dagger}\Phi\Phi^\dagger_E\Phi+\Phi^\dagger\Phi_E\Phi^\dagger\Phi_E)-
\xi_3\mathrm{Tr}(\Phi^{\dagger}\Phi_E\Phi^\dagger_E\Phi_E+\Phi\Phi_E^\dagger\Phi_E\Phi^\dagger_E)\nonumber\\& -
\xi_4\mathrm{Tr}(\Phi^{\dagger}_E\Phi_E
)^2
+\mathrm{Tr}%
(\Phi_E^{\dagger}\Phi_E E_1+\Phi_E\Phi_E^\dagger E_1)+ +c_{1}^*[(\mathrm{det}\Phi-\mathrm{det}\Phi^{\dagger}_E)^{2}+(\mathrm{det}\Phi^\dagger-\mathrm{det}\Phi_E)^{2}]\nonumber\\& +c_{1E}^*(\mathrm{det}\Phi_E-\mathrm{det}\Phi^{\dagger}_E)^{2}
+\frac{h_{1}^*}{2}\mathrm{Tr}(\Phi^{\dagger}_E\Phi+\Phi^\dagger\Phi_E)\mathrm{Tr}\left(  L_{\mu
}^{2}+R_{\mu}^{2}\right)+ \frac{h_{1E}^*}{2}\mathrm{Tr}(\Phi^{\dagger}_E\Phi_E)\mathrm{Tr}\left(  L_{\mu
}^{2}+R_{\mu}^{2}\right) \nonumber\\&  
+h_{2}^*\mathrm{Tr}(\Phi_E^\dagger L_{\mu} L^\mu\Phi+\Phi^\dagger L_\mu L^\mu \Phi_E+R_\mu\Phi_E^\dagger
\Phi R^\mu+ R_{\mu}\Phi^\dagger \Phi_E R^\mu)+h_{2E}^*\mathrm{Tr}[\left\vert L_{\mu}\Phi_E\right\vert
^{2}+\left\vert \Phi_E R_{\mu}\right\vert ^{2}]\nonumber\\
&  +2h_{3}^*\mathrm{Tr}(L_{\mu}\Phi_E R^{\mu}\Phi^{\dagger}+L_\mu\Phi R^\mu \Phi_E^\dagger)+2h_{3E}^*\mathrm{Tr}(L_{\mu}\Phi_E R^{\mu}\Phi^{\dagger}_E)\,. \label{fulllag}%
\end{align}
where  $E_1=diag\{0,0,\epsilon_S^E\}$,
\begin{equation}\label{4}
L^\mu=(V^a+i\,A^a)^{\mu}\,t^a=\frac{1}{\sqrt{2}}
\left(%
\begin{array}{ccc}
  \frac{\omega_N+\rho^{0}}{\sqrt{2}}+ \frac{f_{1N}+a_1^{0}}{\sqrt{2}} & \rho^{+}+a^{+}_1 & K^{*+}+K^{+}_1 \\
  \rho^{-}+ a^{-}_1 &  \frac{\omega_N-\rho^{0}}{\sqrt{2}}+ \frac{f_{1N}-a_1^{0}}{\sqrt{2}} & K^{*0}+K^{0}_1 \\
  K^{*-}+K^{-}_1 & \overline{K}^{*0}+\overline{K}^{0}_1 & \omega_{S}+f_{1S} \\
\end{array}%
\right)^\mu\,,
\end{equation}
and
\begin{equation}\label{5}
R^\mu=(V^a-i\,A^a)^\mu\,t^a=\frac{1}{\sqrt{2}}
\left(%
\begin{array}{ccc}
  \frac{\omega_N+\rho^{0}}{\sqrt{2}}- \frac{f_{1N}+a_1^{0}}{\sqrt{2}} & \rho^{+}-a^{+}_1 & K^{*+}-K^{+}_1\\
  \rho^{-}- a^{-}_1 &  \frac{\omega_N-\rho^{0}}{\sqrt{2}}-\frac{f_{1N}-a_1^{0}}{\sqrt{2}} & K^{*0}-K^{0}_1 \\
  K^{*-}-K^{-}_1 & \overline{K}^{*0}-\overline{K}^{0}_1 & \omega_{S}-f_{1S} \\
\end{array}%
\right)^\mu\,.
\end{equation}
The vector and axial-vector fields $\omega_N,\,\omega_S,\, \overrightarrow{\rho},\,f_{1N},\, f_{1S}, \overrightarrow{a_1},\, K^*, K^*_0$ and $K_1$ are assigned to light physical resonanceses $\omega(782),\,\phi(1020),\,\rho(770),\,f_1(1285),\,f_1(1420),\,a_1(1260),\,K^*(982),\,K^*_0(1430)$ and $K_1(1270)$, respectively. For more details see Ref.\ \cite{Excited-dick}.

The explicit expressions of the wave-function renormalization
constants $Z_{i}$ introduced in Eq.\ (\ref{psz})
read \cite{dick}:%
\begin{equation}
Z_{\pi}=Z_{\eta_{N}}=\frac{m_{a_{1}}}{\sqrt{m_{a_{1}}^{2}-g_{1}^{2}\phi
_{N}^{2}}}\;,\,\, \,\,\,\, \, Z_{K}=\frac{2m_{K_{1}}}{\sqrt{4m_{K_{1}}^{2}-g_{1}^{2}(\phi_{N}+\sqrt{2}%
\phi_{S})^{2}}}\;,\label{zpi}%
\end{equation}%

\begin{equation}
Z_{K_{S}}=\frac{2m_{K^{\star}}}{\sqrt{4m_{K^{\star}}^{2}-g_{1}^{2}(\phi
_{N}-\sqrt{2}\phi_{S})^{2}}}\;,\,\, \,\,\,\, \, 
Z_{\eta_{S}}=\frac{m_{f_{1S}}}{\sqrt{m_{f_{1S}}^{2}-2g_{1}^{2}\phi_{S}^{2}}%
}\;. \label{zets}%
\end{equation}

\subsection{Explicit form of the Lagrangian in Eq.\ (\ref{lag1})}

\label{app2}

After performing the field transformations in Eq.
(\ref{psz}), the effective Lagrangian (\ref{lag1}) takes the form:

\begin{align}\label{exL2a}
\mathcal{L}_{\tilde{G}\Phi\Phi_E}^{int} &  =c_{\tilde{G}\Phi\Phi_E}
\tilde{G}\big\{-\frac{1}{16} \phi_N^4 \big[Z^2_{\eta_S}(\eta_S+\eta_{SE})^2-(\sigma_S-\sigma_{SE})^2]-\frac{1}{4} \phi_N^2 \phi_S^2 [Z^2_{\eta_N}(\eta_N+\eta_{NE})^2-2(\sigma_N-\sigma_{NE})^2\big]
\\ \nonumber &
+\frac{1}{8}\phi_N^3\phi_S\big[2\sqrt{2}\sigma_N\sigma_S-2\sqrt{2}\sigma_{NE}\sigma_S-2\sqrt{2}\sigma_N\sigma_{SE}-2 Z_{\eta_N} Z_{\eta_S} (\eta_N\eta_S+\eta_{NE}\eta_S+\eta_N\eta_{SE}+\eta_{NE}\eta_{SE})\big]
\\ \nonumber &
+\frac{1}{4} \phi_N \phi_S^2\big[2Z_\pi Z_{\eta_N}(a_0^0 \pi^0+a_0^+\pi^-+a_0^-\pi^+)(\eta_N+\eta_{NE})+2Z_\pi Z_{\eta_N}(a_{0E}^0 \pi^0_E+a_{0E}^+\pi^-_E+a_{0E}^-\pi^+_E)\eta_N \big]
\\ \nonumber &
+\frac{1}{\sqrt{2}} \phi_N \phi_S^2\sigma_N\big[(a_0^0)^2+2a_0^-a_0^++Z_\pi^2(\pi^0\pi^0+2\pi^-\pi^+)+Z_{\eta_N}^2(\eta_{NE}^2+2\eta_N\eta_{NE})+Z_\pi^2(\pi^0_E\pi^0_E+2\pi^-_E\pi^+_E)-3Z^2_{\eta_N}\eta_N^2\big]
\\ \nonumber &
-\frac{1}{\sqrt{2}} \phi_N \phi_S^2\sigma_{NE}\big[Z^2_\pi(\pi^0\pi^0+2\pi^+\pi^-+Z^2_{\eta_N}\eta_N^2+2\eta_N\eta_{NE}\big]+\sqrt{2} \phi_N \phi_S^2\sigma_{N}^3+\frac{1}{8}\phi_N^2\phi_SZ_\pi^2(\pi^0\pi^0+2\pi^-\pi^+)(\sigma_S+\sigma_{SE})
\\ \nonumber &
+\frac{1}{2\sqrt{2}}\phi_N^2\phi_S Z_K Z_{K_S}Z_{\eta_N}(K^+K^-_S+K^-K^+_S+K^0\overline{K}^0_S+\overline{K}^0K^0_S)(\eta_{NE}+\eta_N)+\frac{5}{8}\phi_N^2\phi_S\sigma_S(2\sigma_N^2+Z_{\eta_N}^2\eta_N^2)\\ \nonumber &
 +\frac{1}{2}\phi_N^2\phi_S(\sigma_N+\sigma_{NE})\big[Z_K^2(K^-K^++K^0\overline{K}^0)+Z_{K_S}^2(K_S^-K_S^++K_s^0\overline{K}^0_S)\big]-\frac{5}{2\sqrt{2}}\phi_N^2\phi_SZ_{\eta_N}Z_{\eta_S}\eta_N\eta_S\sigma_N \\ \nonumber &
+\frac{1}{8}\phi_N^2\phi_S\,Z^2_{\eta_N}(\eta_N^2\sigma_{SE}+\eta_N\eta_{NE}\sigma_S)+\frac{1}{4}\phi_N^2\phi_S\, Z_\pi Z_{\eta_S}(a_0^0\pi^0+a^+_0\pi^-+a_0^-\pi^+)(\eta_{SE}+\eta_S)
 \\ \nonumber &
-\frac{1}{\sqrt{2}} \phi_N^2\phi_S\,Z_{\eta_N}\,Z_{\eta_S}\big[(\eta_{NE}\eta_S+\frac{1}{2}\eta_N\eta_{SE})\sigma_N+(\eta_N\eta_S-\eta_{NE}\eta_S)\sigma_{NE}\big]\big\}\text{ .}%
\end{align}

Note that, several decay channels of the pseudoscalar glueball, $\tilde{G}$, are not kinematically allowed, because the summation mass of the decay products is larger than the mass of the decaying particles $M<\sum^3_i\,\, m_i$, which appear in Eq. ($\ref{exL2a}$) and are not present in Table I.

\subsection{Explicit form of the Lagrangian in Eq.\ (\ref{intlag3})}

\label{app4}

From Eq.(\ref{intlag3}), we obtain the following corresponding interaction Lagrangian by developing the field transformations in Eq. (\ref{psz}) 
\begin{align}\label{exL3}
\mathcal{L}_{\tilde{G}\Phi\Phi_E}^{int} &  =c_{\tilde{G}\Phi\Phi_E}
\tilde{G}^*\big\{-\frac{1}{16} \phi_N^4 \big[Z^2_{\eta_S}(\eta_S+\eta_{SE})^2-(\sigma_S-\sigma_{SE})^2]-\frac{1}{4} \phi_N^2 \phi_S^2 [Z^2_{\eta_N}(\eta_N+\eta_{NE})^2-2(\sigma_N-\sigma_{NE})^2\big]
\\ \nonumber &
+\frac{1}{8}\phi_N^3\phi_S\big[2\sqrt{2}\sigma_N\sigma_S-2\sqrt{2}\sigma_{NE}\sigma_S-2\sqrt{2}\sigma_N\sigma_{SE}-2 Z_{\eta_N} Z_{\eta_S} (\eta_N\eta_S+\eta_{NE}\eta_S+\eta_N\eta_{SE}+\eta_{NE}\eta_{SE})\big]
\\ \nonumber &
+\frac{1}{4} \phi_N \phi_S^2\big[2Z_\pi Z_{\eta_N}(a_0^0 \pi^0+a_0^+\pi^-+a_0^-\pi^+)(\eta_N+\eta_{NE})+2Z_\pi Z_{\eta_N}(a_{0E}^0 \pi^0_E+a_{0E}^+\pi^-_E+a_{0E}^-\pi^+_E)\eta_N \big]
\\ \nonumber &
+\frac{1}{\sqrt{2}} \phi_N \phi_S^2\sigma_N\big[(a_0^0)^2+2a_0^-a_0^++Z_\pi^2(\pi^0\pi^0+2\pi^-\pi^+)+Z_{\eta_N}^2(\eta_{NE}^2+2\eta_N\eta_{NE})+Z_\pi^2(\pi^0_E\pi^0_E+2\pi^-_E\pi^+_E)-3Z^2_{\eta_N}\eta_N^2\big]
\\ \nonumber &
-\frac{1}{\sqrt{2}} \phi_N \phi_S^2\sigma_{NE}\big[Z^2_\pi(\pi^0\pi^0+2\pi^+\pi^-+Z^2_{\eta_N}\eta_N^2+2\eta_N\eta_{NE}\big]+\sqrt{2} \phi_N \phi_S^2\sigma_{N}^3+\frac{1}{8}\phi_N^2\phi_SZ_\pi^2(\pi^0\pi^0+2\pi^-\pi^+)(\sigma_S+\sigma_{SE})
\\ \nonumber &
+\frac{1}{2\sqrt{2}}\phi_N^2\phi_S Z_K Z_{K_S}Z_{\eta_N}(K^+K^-_S+K^-K^+_S+K^0\overline{K}^0_S+\overline{K}^0K^0_S)(\eta_{NE}+\eta_N)+\frac{5}{8}\phi_N^2\phi_S\sigma_S(2\sigma_N^2+Z_{\eta_N}^2\eta_N^2)\\ \nonumber &
 +\frac{1}{2}\phi_N^2\phi_S(\sigma_N+\sigma_{NE})\big[Z_K^2(K^-K^++K^0\overline{K}^0)+Z_{K_S}^2(K_S^-K_S^++K_s^0\overline{K}^0_S)\big]-\frac{5}{2\sqrt{2}}\phi_N^2\phi_SZ_{\eta_N}Z_{\eta_S}\eta_N\eta_S\sigma_N \\ \nonumber &
+\frac{1}{8}\phi_N^2\phi_S\,Z^2_{\eta_N}(\eta_N^2\sigma_{SE}+\eta_N\eta_{NE}\sigma_S)+\frac{1}{4}\phi_N^2\phi_S\, Z_\pi Z_{\eta_S}(a_0^0\pi^0+a^+_0\pi^-+a_0^-\pi^+)(\eta_{SE}+\eta_S)
 \\ \nonumber &
-\frac{1}{\sqrt{2}} \phi_N^2\phi_S\,Z_{\eta_N}\,Z_{\eta_S}\big[(\eta_{NE}\eta_S+\frac{1}{2}\eta_N\eta_{SE})\sigma_N+(\eta_N\eta_S-\eta_{NE}\eta_S)\sigma_{NE}\big]\big\}\text{ .}%
\end{align}
For the particles reported in Tables II, III, IV, and V.
\subsection{Explicit form of the Lagrangian in Eq.\ (\ref{intlag2})}

\label{app3}

After applying the field transformations in Eq.
(\ref{psz}), the chiral effective Lagrangian (\ref{intlag2}) takes the form:

\begin{align}\label{exL2}
\mathcal{L}_{\tilde{G}\Phi}  &  =\frac{1}{2\sqrt{2}}c_{\tilde{G}^*\Phi_E}
\tilde{G}^*\big[ 2Z_\pi(a^0_{0E}\pi^0_E+a^+_{0E}\pi^-_E+a^-_{0E}\pi^+_E)\phi_S\\ \nonumber &
+\sqrt{2}Z_K\,Z_{K_S}\phi_{NE}(K^-_EK_{SE}^++K^0_E\overline{K}^0_{SE}+K^0_{SE}\overline{K}^0_E+K^-_{SE}K^+_E)\big] \\ \nonumber &
-2Z_{\eta_S}\phi_N\eta_{SE}\sigma_{NE}-2Z_{\eta_N}\phi_N\eta_{NE}\sigma_{SE}-2Z_{\eta_N}\eta_{NE}\sigma_{NE}\Phi_S\text{ .}%
\end{align}

\subsection{Two-body decay}

The general formula of the two-body
decay width \cite{EshraimTH} is given by

\begin{equation}
\label{B1}\Gamma_{P\rightarrow P_1 P_2}=\frac{S_{P\rightarrow P_1 P_2}k(m_{P}%
,\,m_{P_1},\,m_{P_2})}{8 \pi m_{P}^{2}}|\mathcal{M}_{P\rightarrow
P_1 P_2}|^{2},
\end{equation}

where P is the decaying particle, $P_1$ and $P_2$ are the decay products, $k(m_{P}%
,\,m_{P_1},\,m_{P_2})$ is the center-of-mass momentum of $P_1$ and $P_2$ and described as 
\begin{equation}
k(m_{P},\,m_{P_1},\,m_{P_2})=\frac{1}{2m_{P}}\sqrt{m_{P}^{4}+(m_{P_1}^{2}-m_{P_2}%
^{2})^{2}-2m_{P}^{2}\,(m_{P_1}^{2}+m_{P_2}^{2})}\theta(m_{P}-m_{P_1}-m_{P_2})\,,
\label{B2}%
\end{equation}
$\mathcal{M}_{P\rightarrow P_1 P_2}$ refers to the corresponding tree-level
decay amplitude, and $S_{P\rightarrow P_1 P_2}$ is a
symmetrization factor (it equals $1/2$ for two identical particles in the final state and it equals $1$ if $P_1$ and $P_2$ are different). The $\theta$ function ensures that the mass of the particles produced in the decay does not exceed the initial mass.

\subsection{Three-body decay}

\label{app3}

The general explicit expression for the three-body decay
width for the process $P\rightarrow P_{1}P_{2}P_{3}$ \cite{PDG}:%
\[
\Gamma_{P\rightarrow P_{1}P_{2}P_{3}}=\frac{s_{P\rightarrow
P_{1}P_{2}P_{3}}}{32(2\pi)^{3}M_{P}^{3}}\int_{(m_{1}+m_{2})^{2}%
}^{(M_{P}-m_{3})^{2}}dm_{12}^{2}\int_{(m_{23})_{\min}}^{(m_{23}%
)_{\max}}|-i\mathcal{M}_{P\rightarrow P_{1}P_{2}P_{3}}|^{2}dm_{23}^{2}%
\]
where
\begin{align}
(m_{23})_{\min}  &  =(E_{2}^{\ast}+E_{3}^{\ast})^{2}-\left(  \sqrt{E_{2}%
^{\ast2}-m_{2}^{2}}+\sqrt{E_{3}^{\ast2}-m_{3}^{2}}\right)  ^{2}\text{ ,}\\
(m_{23})_{\max}  &  =(E_{2}^{\ast}+E_{3}^{\ast})^{2}-\left(  \sqrt{E_{2}%
^{\ast2}-m_{2}^{2}}-\sqrt{E_{3}^{\ast2}-m_{3}^{2}}\right)  ^{2}\text{ ,}%
\end{align}
and%
\begin{equation}
E_{2}^{\ast}=\frac{m_{12}^{2}-m_{1}^{2}+m_{2}^{2}}{2m_{12}}\text{ , }%
E_{3}^{\ast}=\frac{M_{P}^{2}-m_{12}^{2}-m_{3}^{2}}{2m_{12}}\text{ .}%
\end{equation}
The quantities $m_{1},$ $m_{2},$ $m_{3}$ refer to
the masses of the three
decay products $P_{1},$ $P_{2}$, and $P_{3}$,  $\mathcal{M}_{
P\rightarrow P_{1}P_{2}P_{3}}$ denotes the decay amplitude of the tree-level, and the symmetrization
factor $s_{P\rightarrow P_{1}P_{2}P_{3}}$ equals $6$ when $P_{1},$ $P_{2}$, and $P_{3}$ are different, equals $2$ when two of the particles are identical in the final state, and equals $1$ when the three decay products are identical.


\begin{thebibliography}{99}                                                                                               %





\bibitem {bag-glueball}
R.~L.~Jaffe and K.~Johnson,
Phys.\ Lett.\ B \textbf{60}, 201 (1976);
R.~Konoplich and M.~Shchepkin,
Nuovo Cim.\ A \textbf{67}, 211 (1982);
M.~Jezabek and J.~Szwed,
Acta Phys.\ Polon.\ B \textbf{14}, 599 (1983);
R.~L.~Jaffe, K.~Johnson and Z.~Ryzak,
Annals Phys.\ \textbf{168}, 344 (1986);
M.~Strohmeier-Presicek, T.~Gutsche, R.~Vinh Mau and A.~Faessler,
Phys.\ Rev.\ D \textbf{60}, 054010 (1999) [arXiv:hep-ph/9904461].



\bibitem{Nakano}
T. Nakano, et al., Phys. Rev. Lett. 91, 012002 (2003).




\bibitem {Morningstar}C.~Morningstar and M.~J.~Peardon,
AIP Conf.\ Proc.\ \textbf{688}, 220 (2004) [arXiv:nucl-th/0309068];C.~J.~Morningstar and M.~J.~Peardon,
  Phys.\ Rev.\ D {\bf 60}, 034509 (1999)
  [hep-lat/9901004];
 M. Strohmeier-Presicek, T. Gutsche, R. Vinh Ma
u and A. Faessler, Phys. Rev.\ D {\bf
60}, 054010 (1999)
[arXiv:hep-ph/9904461]
M.~Loan, X.~Q.~Luo and Z.~H.~Luo,
Int.\ J.\ Mod.\ Phys.\ A \textbf{21}, 2905 (2006) [arXiv:hep-lat/0503038];
E.~B.~Gregory, A.~C.~Irving, C.~C.~McNeile, S.~Miller and Z.~Sroczynski,
PoS \textbf{LAT2005}, 027 (2006) [arXiv:hep-lat/0510066].
\bibitem{Chen}
Y.~Chen \textit{et al.},
Phys.\ Rev.\ D \textbf{73}, 014516 (2006) [arXiv:hep-lat/0510074].

\bibitem {PDG}K. Nakamura \textit{et al}. (Particle Data Group), J. Phys. G
\textbf{37}, 075021 (2010).
\bibitem {bes}M. Ablikim \textit{et al.} (BES Collaboration), Phys. Rev. Lett.
\textbf{95}, 262001 (2005); N. Kochelev and D. P. Min, Phys. Lett. B
\textbf{633}, 283 (2006); M. Ablikim \textit{et al.} (BES III Collaboration),
Phys. Rev. Lett. 106.072002 (2011).


\bibitem {panda}M.~F.~M.~Lutz \textit{et al.} [ PANDA Collaboration ],
arXiv:0903.3905 [hep-ex]].

\bibitem{NICA} 
  D.~Parganlija,
  Eur.\ Phys.\ J.\ A {\bf 52}, no. 8, 229 (2016)
  [arXiv:1601.05328 [hep-ph]].


\bibitem {review}F.~E.~Close,
Rept.\ Prog.\ Phys.\ \textbf{51}, 833 (1988);
S.~Godfrey and J.~Napolitano,
Rev.\ Mod.\ Phys.\ \textbf{71}, 1411 (1999) [arXiv:hep-ph/9811410];
C.~Amsler and N.~A.~Tornqvist,
Phys.\ Rept.\ \textbf{389}, 61 (2004);
E.~Klempt and A.~Zaitsev,
Phys.\ Rept.\ \textbf{454}, 1 (2007) [arXiv:0708.4016 [hep-ph]].
\bibitem{Ochs}
W.~Ochs,
  J.\ Phys.\ G {\bf 40}, 043001 (2013)
  [arXiv:1301.5183 [hep-ph]]; V.~Mathieu, N.~Kochelev and V.~Vento,
  Int.\ J.\ Mod.\ Phys.\ E {\bf 18}, 1 (2009)
  [arXiv:0810.4453 [hep-ph]];
   V.~Crede and C.~A.~Meyer,
  Prog.\ Part.\ Nucl.\ Phys.\  {\bf 63}, 74 (2009)
  [arXiv:0812.0600 [hep-ex]].

\bibitem{Sonnenschein} 
 J.~Sonnenschein and D.~Weissman,
  Eur.\ Phys.\ J.\ C {\bf 79}, no. 4, 326 (2019)
  [arXiv:1812.01619 [hep-ph]].
\bibitem{Novikov}
 V.~A.~Novikov, M.~A.~Shifman, A.~I.~Vainshtein and V.~I.~Zakharov,
  Nucl.\ Phys.\ B {\bf 165}, 55 (1980);  V.~A.~Novikov, M.~A.~Shifman, A.~I.~Vainshtein and V.~I.~Zakharov,
  Nucl.\ Phys.\ B {\bf 165}, 67 (1980).

\bibitem{Pimikov}  
 A.~Pimikov, H.~J.~Lee and N.~Kochelev,
  Phys.\ Rev.\ Lett.\  {\bf 119}, no. 7, 079101 (2017)
  [arXiv:1702.06634 [hep-ph]];  A.~Pimikov, H.~J.~Lee, N.~Kochelev, P.~Zhang and V.~Khandramai,
  Phys.\ Rev.\ D {\bf 96}, no. 11, 114024 (2017)
 [arXiv:1708.07675 [hep-ph]].

\bibitem{Masoni} 
 A.~Masoni, C.~Cicalo and G.~L.~Usai,
  J.\ Phys.\ G {\bf 32}, R293 (2006); V.~Mathieu and V.~Vento,
  Phys.\ Rev.\ D {\bf 81}, 034004 (2010)
  [arXiv:0910.0212 [hep-ph]].
  

\bibitem{EshraimG}W. I. Eshraim, S. Janowski, F. Giacosa and D. H. Rischke,
Phys. Rev. D \textbf{87}, 054036 (2013) [arXiv:1208.6474
[hep-ph]].
\bibitem{WGNuclion}
W. I. Eshraim, S. Janowski, A. Peters, K. Neuschwander
and F. Giacosa, Acta Phys. Polon. Supp. \textbf{5}, 1101 (2012)
[arXiv:1209.3976 [hep-ph]].
\bibitem{Volkov}
M.~K.~Volkov, V.~L.~Yudichev and M.~Nagy,
  Nuovo Cim.\ A {\bf 112}, 955 (1999).
  
\bibitem{Eshraim-schramm} 
  W.~I.~Eshraim and S.~Schramm,
  Phys.\ Rev.\ D {\bf 95}, no. 1, 014028 (2017)
  [arXiv:1606.02207 [hep-ph]].

\bibitem{Brunner} 
  F.~Brünner and A.~Rebhan,
  Phys.\ Lett.\ B {\bf 770}, 124 (2017)
  [arXiv:1610.10034 [hep-ph]].  
\bibitem{Rosenzweig}  
  C.~Rosenzweig, A.~Salomone and J.~Schechter,
  Phys.\ Rev.\ D {\bf 24}, 2545 (1981);  C.~Rosenzweig, A.~Salomone and J.~Schechter,
  Nucl.\ Phys.\ B {\bf 206}, 12 (1982),
  Erratum: [Nucl.\ Phys.\ B {\bf 207}, 546 (1982)]; C.~Rosenzweig, J.~Schechter and C.~G.~Trahern,
  Phys.\ Rev.\ D {\bf 21}, 3388 (1980).
  
\bibitem{Ohta}  
  K.~Kawarabayashi and N.~Ohta,
  Nucl.\ Phys.\ B {\bf 175}, 477 (1980); K.~Kawarabayashi and N.~Ohta,
  Prog.\ Theor.\ Phys.\  {\bf 66}, 1789 (1981).


\bibitem{Lattice} B. Berg and A. Billoire, Nucl. Phys. \textbf{B221}, 109 (1983); G. Bali, \textit{et al.} (UKQCD Collaboration), Phys. Lett. \textbf{B309}, 378 (1993); C. Michael and M. Teper, Nucl. Phys. \textbf{B314}, 347 (1989). C. Morningstar and M. Peardon, Phys. Rev. \textbf{D56}, 4043 (1997); Phys. Rev. \textbf{D60}, 034509 (1999).

\bibitem{ExLattice1}
 P.~Lacock {\it et al.} [UKQCD Collaboration],
  Phys.\ Rev.\ D {\bf 54}, 6997 (1996)
  [hep-lat/9605025]; T.~Burch, C.~Gattringer, L.~Y.~Glozman, C.~Hagen, C.~B.~Lang and A.~Schafer,
  Phys.\ Rev.\ D {\bf 73}, 094505 (2006)
  [hep-lat/0601026].
\bibitem{ExLattice2}
 J.~J.~Dudek, R.~G.~Edwards, M.~J.~Peardon, D.~G.~Richards and C.~E.~Thomas,
  Phys.\ Rev.\ Lett.\  {\bf 103}, 262001 (2009)
  [arXiv:0909.0200 [hep-ph]]; J.~J.~Dudek {\it et al.} [Hadron Spectrum Collaboration],
  Phys.\ Rev.\ D {\bf 88}, no. 9, 094505 (2013)
  [arXiv:1309.2608 [hep-lat]].
\bibitem{Badalian}
 A.~M.~Badalian and B.~L.~G.~Bakker,
  Phys.\ Rev.\ D {\bf 66}, 034025 (2002)
  [hep-ph/0202246].
\bibitem{NJL}
M.~K.~Volkov, D.~Ebert and M.~Nagy,
  Int.\ J.\ Mod.\ Phys.\ A {\bf 13}, 5443 (1998)
  [hep-ph/9705334]; M.~K.~Volkov, V.~L.~Yudichev and D.~Ebert,
  J.\ Phys.\ G {\bf 25}, 2025 (1999)
  [JINR Rapid Commun.\  {\bf 6-92}, 5 (1998)]
  [hep-ph/9810470]; M.~K.~Volkov and V.~L.~Yudichev,
  Int.\ J.\ Mod.\ Phys.\ A {\bf 14}, 4621 (1999)
  [hep-ph/9904226]; M.~K.~Volkov and V.~L.~Yudichev,
  Phys.\ Atom.\ Nucl.\  {\bf 63}, 455 (2000)
  [Yad.\ Fiz.\  {\bf 63}, 527 (2000)];  A.~V.~Vishneva and M.~K.~Volkov,
  Int.\ J.\ Mod.\ Phys.\ A {\bf 29}, no. 24, 1450125 (2014)
  [arXiv:1403.1360 [hep-ph]].
\bibitem{Bethe}
 A.~Holl, A.~Krassnigg and C.~D.~Roberts,
  Phys.\ Rev.\ C {\bf 70}, 042203 (2004)
  [nucl-th/0406030]; A.~Holl, A.~Krassnigg, C.~D.~Roberts and S.~V.~Wright,
  Int.\ J.\ Mod.\ Phys.\ A {\bf 20}, 1778 (2005)
  [nucl-th/0411065]; B.~L.~Li, L.~Chang, F.~Gao, C.~D.~Roberts, S.~M.~Schmidt and H.~S.~Zong,
  Phys.\ Rev.\ D {\bf 93}, no. 11, 114033 (2016)
  [arXiv:1604.07415 [nucl-th]].


\bibitem{excited-mesons} 
  T.~Ino,
  Prog.\ Theor.\ Phys.\  {\bf 71}, 864 (1984);  P.~Geiger,
  Phys.\ Rev.\ D {\bf 49}, 6003 (1994)
  [hep-ph/9311254]; S.~M.~Fedorov and Y.~A.~Simonov,
  JETP Lett.\  {\bf 78}, 57 (2003)
  [Pisma Zh.\ Eksp.\ Teor.\ Fiz.\  {\bf 78}, 67 (2003)].
  
\bibitem{excited-mesons2}  
  T.~Gutsche, V.~E.~Lyubovitskij and M.~C.~Tichy,
  Phys.\ Rev.\ D {\bf 79}, 014036 (2009)
  [arXiv:0811.0668 [hep-ph]]; G.~Rupp, S.~Coito and E.~van Beveren,
  Acta Phys.\ Polon.\ Supp.\  {\bf 9}, 653 (2016)
  [arXiv:1605.04260 [hep-ph]].
  
\bibitem{Excited-dick} 
  D.~Parganlija and F.~Giacosa,
  Eur.\ Phys.\ J.\ C {\bf 77}, no. 7, 450 (2017)
  [arXiv:1612.09218 [hep-ph]]. 
\bibitem{DickNF2} 
  D.~Parganlija, F.~Giacosa and D.~H.~Rischke,
  Phys.\ Rev.\ D {\bf 82}, 054024 (2010)
  [arXiv:1003.4934 [hep-ph]].
  

\bibitem {dick}
  D.~Parganlija, P.~Kovacs, G.~Wolf, F.~Giacosa and D.~H.~Rischke,
  Phys.\ Rev.\ D {\bf 87}, no. 1, 014011 (2013)
  [arXiv:1208.0585 [hep-ph]].
 
\bibitem{staniglueball} 
  S.~Janowski, F.~Giacosa and D.~H.~Rischke,
  Phys.\ Rev.\ D {\bf 90}, no. 11, 114005 (2014)
  [arXiv:1408.4921 [hep-ph]].
  
\bibitem{Bary}
  S.~Gallas, F.~Giacosa and D.~H.~Rischke,
  Phys.\ Rev.\ D {\bf 82}, 014004 (2010)
  [arXiv:0907.5084 [hep-ph]].
  
\bibitem{wcharm1}W.~I.~Eshraim, F.~Giacosa and D.~H.~Rischke,
  Eur.\ Phys.\ J.\ A {\bf 51}, no. 9, 112 (2015)
  [arXiv:1405.5861 [hep-ph]]; W.~I.~Eshraim,
  PoS QCD {\bf -TNT-III}, 049 (2013)
  [arXiv:1401.3260 [hep-ph]]; W.~I.~Eshraim and F.~Giacosa,
  EPJ Web Conf.\  {\bf 81}, 05009 (2014)
  [arXiv:1409.5082 [hep-ph]]; W.~I.~Eshraim,
  EPJ Web Conf.\  {\bf 95}, 04018 (2015)
  [arXiv:1411.2218 [hep-ph]]; W.~I.~Eshraim,
  J.\ Phys.\ Conf.\ Ser.\  {\bf 599}, no. 1, 012009 (2015)
  [arXiv:1411.4749 [hep-ph]].

\bibitem{EshraimTH} 
  W.~I.~Eshraim, Ph.D. thesis, Frankfurt U.(2015)
  [arXiv:1509.09117 [hep-ph]].

\bibitem{wcharm2}
  W.~I.~Eshraim and C.~S.~Fischer,
  Eur.\ Phys.\ J.\ A {\bf 54}, no. 8, 139 (2018)
  [arXiv:1802.05855 [hep-ph]]; W.~I.~Eshraim,
  EPJ Web Conf.\  {\bf 126}, 04017 (2016).






\end{thebibliography}
\end{document}